\documentclass[journal=jpccck,manuscript=article,maxauthors=10]{achemso}

\setcitestyle{super}         

\usepackage[utf8]{inputenc}
\usepackage[T1]{fontenc}
\usepackage[english]{babel}
\usepackage{graphicx}
\usepackage{subcaption}
\usepackage{ragged2e}
\usepackage{amsmath}
\usepackage{bigints}
\usepackage{changes}
\usepackage{xcolor}
\usepackage{hyperref}
\hypersetup{colorlinks, allcolors=black}
\usepackage{todonotes}
\usepackage{ragged2e}
\usepackage{placeins}

\definecolor{addedcolor}{rgb}{0,0.5,0}
\definecolor{deletedcolor}{rgb}{1,0,0}
\setdeletedmarkup{\textcolor{deletedcolor}{\sout{#1}}}
\setaddedmarkup{\textcolor{addedcolor}{#1}}

\DeclareUnicodeCharacter{2212}{\ensuremath{-}}

\title{Cross-polarized and Stable Second Harmonic Generation from Monocrystalline Copper}
\author{Elif Nur Dayi}
\author{Alan R. Bowman}
\author{Omer Can Karaman}
\author{Diotime Pellet}
\author{Giulia Tagliabue}
\email{giulia.tagliabue@epfl.ch}
\affiliation{Laboratory of Nanoscience for Energy Technologies (LNET), STI, \\École Polytechnique Fédérale de Lausanne, 1015 Lausanne, Switzerland}

\begin{document}

\begin{abstract}

Second-harmonic generation (SHG) is a powerful surface-specific probe for centrosymmetric materials, with broad relevance to energy and biological interfaces. Plasmonic nanomaterials have been extensively utilized to amplify this nonlinear response. Yet, material instability has constrained most studies to gold, despite the significance of plasmonic metals such as copper for catalysis. Here, we demonstrate stable and anisotropic SHG from monocrystalline copper, overcoming long-standing challenges associated with surface degradation. By leveraging an on-substrate synthesis approach that yields atomically flat and oxidation-resistant Cu microflakes, we enable reliable SHG measurements and reveal a strong cross-polarized response with \( C_{3v} \) surface symmetry. The SHG signal remains stable over several minutes of continuous femtosecond excitation, highlighting the optical robustness of the Cu microflakes. These results reinforce the viability of monocrystalline Cu as a robust platform for nonlinear nanophotonics and surface-sensitive spectroscopy, expanding the range of copper-based optical applications.

\end{abstract}

\section{Introduction} \label{sec:Intro}
Nonlinear optical characterization has emerged as a powerful approach to probe the symmetry and structural properties of nanomaterials and to control light–matter interactions at the nanoscale \cite{boyd_nonlinear_2020,kauranen_nonlinear_2012}. Among nonlinear processes, second harmonic generation (SHG) is particularly valuable due to its sensitivity to inversion symmetry breaking and consequently its surface-selective nature in centrosymmetric materials\cite{butet_optical_2015}. As a non-destructive technique with broad applicability, SHG is widely employed in plasmonics \cite{chen_interfacial_2024,kauranen_nonlinear_2012}, nanophotonics \cite{stanton_efficient_2020,klimmer_all-optical_2021}, biosensing\cite{tran_applications_2017,aghigh_second_2023}, and heterogeneous catalysis \cite{fromondi_dynamics_2014,nahalka_mapping_2020,pettinger_specific_1992}.

 SHG has been extensively studied in materials including noble metals such as gold \cite{capretti_multipolar_2012,mcnally_nonlinear_2022}, silver \cite{li_second-harmonic_1994,chen_interfacial_2024,chen_surface-enhanced_1981}, as well as in two-dimensional materials \cite{debow_two-dimensional_2024,zhou_strong_2015} and transition metal dichalcogenides \cite{klimmer_all-optical_2021,shi_3r_2017,song_second_2018} in the past decades. However, the nonlinear optical behavior of copper, a material of paramount importance for energy conversion\cite{ramanujam_copper_2017}, catalysis\cite{nitopi_progress_2019,li_electroreduction_2020} and sensing\cite{muniz-miranda_surface-enhanced_2011,pei_disposable_2014}, remains largely unexplored. This is primarily due to its metastable surface chemistry and tendency to oxidation under ambient conditions, which can significantly influence measurements \cite{krause_optical_2004, vollmer_second_1996, baten_second_2007,bloch_detection_1992}.  Moreover, previous investigations have focused almost exclusively on polycrystalline Cu films \cite{petrocelli_wavelength_1993,baten_second_2007,chen_surface-enhanced_1981,krause_optical_2004,lupke_resonant_1994}, where randomly oriented grains average out any anisotropic response and limit insights into the fundamental nature of Cu based SHG behavior.

Direct access to the intrinsic nonlinear optical response of monocrystalline Cu offers a unique opportunity to investigate SHG in centrosymmetric metals beyond gold, thereby expanding the fundamental understanding of second-order processes in plasmonic systems. In this work, leveraging our recently reported oxidation-resistant Cu monocrystalline flakes~\cite{dayi_large-area_2025}, we report reliable SHG measurements from monocrystalline Cu(111) surfaces under ambient conditions. Using power- and polarization-dependent measurements, we demonstrate that Cu microflakes exhibit dominant co- and cross-polarized SHG components, depending on whether the incident polarization is aligned or anti-aligned with the crystal axis. The observed threefold polarization dependence is consistent with \( C_{3v} \) symmetry, as expected for fcc (111) surfaces~\cite{bloch_detection_1992}. In contrast, polycrystalline Cu films display significantly weaker, isotropic emission. These findings highlight the viability of oxidation-resistant monocrystalline Cu microflakes as a robust platform for nonlinear nanophotonics and surface-sensitive spectroscopy, expanding the design space for copper-based optical and catalytic materials.


\section{Methods} \label{sec:Methods}
\subsection*{Sample fabrication}

Cu microflakes were grown using the optimized recipe described in detail in our previous work\cite{dayi_large-area_2025}. Polycrystalline Cu films were sputter-coated on a glass substrate at room temperature (Alliance-Concept DP650). Samples were stored in a nitrogen-purged environment prior to measurement to minimize oxidation.

\subsection*{Second-harmonic generation (SHG) measurements}

SHG measurements were performed using a customized NT$\&$C setup consisting of an inverted microscope (Nikon Eclipse T2) coupled to a spectrometer (Princeton Instruments HRS-500). A 1030 nm femtosecond laser (NKT Origami) was focused onto the sample with a 60$\times$ long working distance objective (NA = 0.7), and the incident power was adjusted using a variable neutral density filter. The emitted SHG signal was first directed by a 750 nm dichroic mirror, which reflects the SHG light (at 515 nm) and transmits the residual fundamental (1030 nm). An 850 nm short-pass filter placed after the dichroic further suppressed any remaining fundamental laser. The filtered SHG signal was then collected through a 4-f lens system before reaching the spectrometer or CCD camera for detection. Measured SHG intensities were normalized by excitation power, illumination area, and integration time to allow comparison between samples.

Power-dependent measurements were acquired by varying the fundamental laser intensity from 0.35 to 2.15~mW/$\mu$m$^2$, and the SHG intensity was fitted on a log–log scale to extract the power-law exponent.

For polarization-resolved SHG measurements, the sample was rotated in \( 5^\circ \)steps using a motorized stage, with 30 seconds of exposure at each condition. The coordinate system was defined with the x- and y-axes lying in the sample plane and the z-axis normal to the surface, while the polarization angle was defined as the in-plane polarization angle ($\theta$) for both emission and excitation. Co- and cross-polarized emission components were collected using a polarizer placed before the spectrometer.

\section{Results \& Discussion } \label{sec:Results}
\begin{figure}[ht!]
    \centering  
\includegraphics[width=10cm]{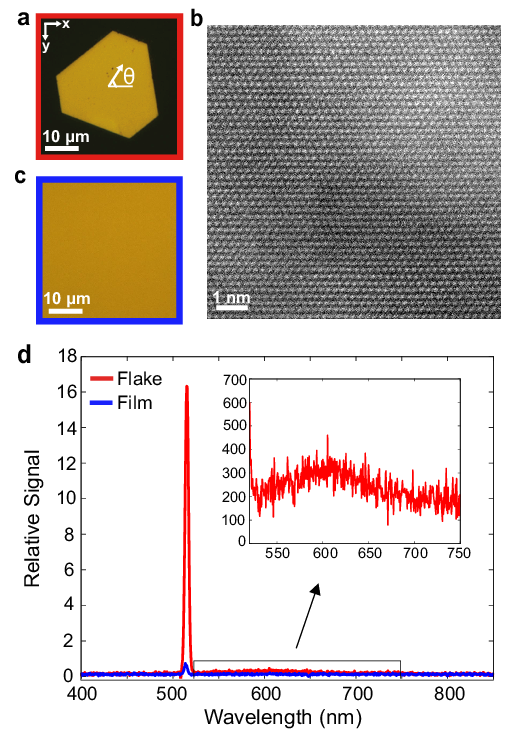}
\caption{\justifying{ \textbf{(a)} Bright-field image of a monocrystalline Cu microflake grown on a glass substrate \textbf{(b)}  High-angle annular dark-field (HAADF) transmission electron microscopy (TEM) image of the Cu microflake revealing the single-crystallinity. \textbf{(c)} Bright-field optical image of a sputtered polycrystalline Cu thin film on glass. \textbf{(d)} Relative SHG intensities for monocrystalline microflakes (red) and polycrystalline films (blue) when exciting at 1030 nm with 2.15  mW/$\mu$m$^2$ beam intensity. Inset shows expanded view of the two-photon photoluminescence response of the microflake in the 550-750 nm range. }}  
\label{fig_flakefilm}
\end{figure}
Monocrystalline Cu microflakes grown with our recently developed on-substrate synthesis method \cite{dayi_large-area_2025} achieve lateral dimensions of  tens of micrometers while advanced structural characterization has shown that they exhibit atomically-flat top surfaces with a well-defined (111) surface orientation. Remarkably, they also display long-term resistance to surface oxidation: no oxide formation is observed even after a month, enabled by the presence of an atomically thin bromide adlayer. Figure \ref{fig_flakefilm}a shows a bright-field image of a studied Cu microflake, its single-crystalline nature clearly recognizable by the truncated triangular shape. The high-angle annular dark-field scanning transmission electron microscopy (HAADF-STEM) image in Figure \ref{fig_flakefilm}b further confirms the monocrystalline face-centered cubic structure of the flake, which plays a key role in determining its nonlinear optical response. Polycrystalline Cu films with a thickness of 100 nm were also prepared, as shown in Figure \ref{fig_flakefilm}c, to enable a direct comparison of the SHG signals from monocrystalline vs. polycrystalline samples.

Under 1030 nm femtosecond laser excitation, the 515 nm SHG signal is much more pronounced for the monocrystalline Cu microflake than the polycrystalline film (Figure \ref{fig_flakefilm}d ). Additionally, the broad feature present in the flake spectrum between 550 nm and 650 nm, as seen in the inset, suggests a weak two-photon luminescence (TPL) arising from the interband and intraband transitions in Cu \cite{wang_cu_2005}.

\begin{figure}[H]
 \centering
\includegraphics[width=9cm]{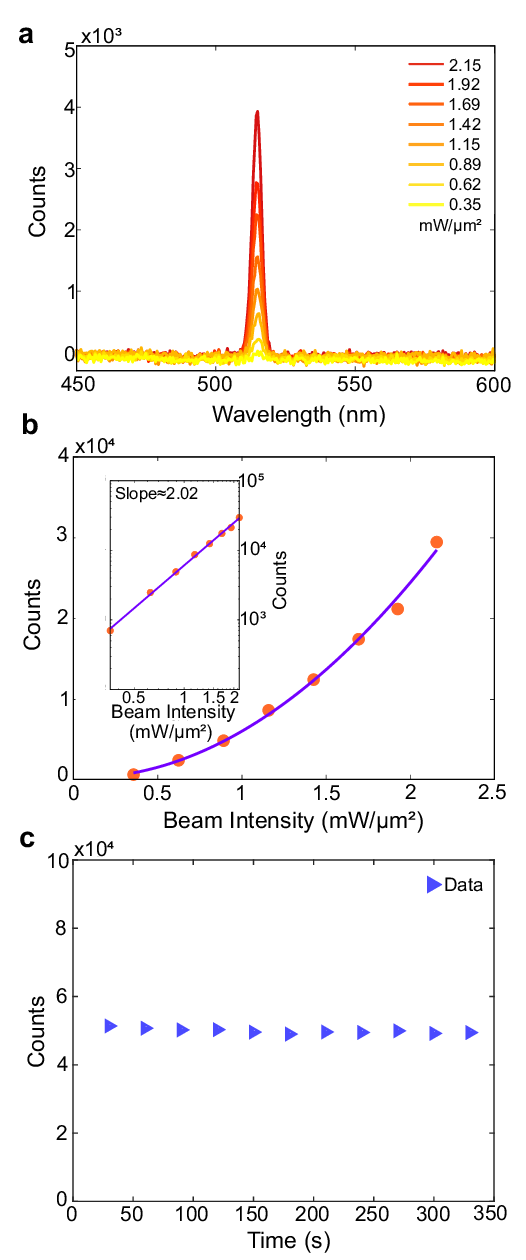}
\caption{\justifying{ \textbf{(a)} SHG intensity from the Cu microflake under varying beam intensities of 0.35 to 2.15 mW/$\mu$m$^2$\textbf{(b)} Quadratic increase of SHG intensity with respect to the powers shown in part a. The inset shows the logarithm curve with a slope of \(2.02 \pm 0.08 \) , which strongly agrees with the theoretical value of 2 expected for second-order processes. \textbf{(c)} Stability test of Cu microflake when irradiated with a beam intensity of 2.15 mW/$\mu$m$^2$ for over 5 minutes. }}
\label{fig_power}
\end{figure}

To confirm the second-order nature of the SHG and ensure the microflake stability under illumination, we performed power-dependent and time-dependent SHG measurements. Figure \ref{fig_power} a and b show the increase in SHG intensity with increasing laser intensity from 0.35 to 2.15  mW/$\mu$m$^2$. The log–log plot in the inset of Figure \ref{fig_power}b yields a slope of $\approx$ \(2.02 \pm 0.08 \), in excellent agreement with the expected value of 2 for a second-order process.

The stability of the Cu flakes under continuous illumination was tested by monitoring the SHG intensity over 5 minutes at a beam intensity of 2.15 mW/$\mu$m$^2$, as shown in Figure \ref{fig_power}c.  Given the surface-sensitive nature of SHG, any degradation due to oxidation or damage would be evident as a change in signal. The nearly constant SHG signal, exhibiting only 1.4\% fluctuation relative to the mean intensity, indicates the absence of visible perturbations from progressive surface oxidation or pump-induced photo-damage. This confirms the optical stability of the microflakes and demonstrates that stable SHG can be obtained from monocrystalline copper.

\begin{figure}[H]
 \centering
 \includegraphics[width=17cm]{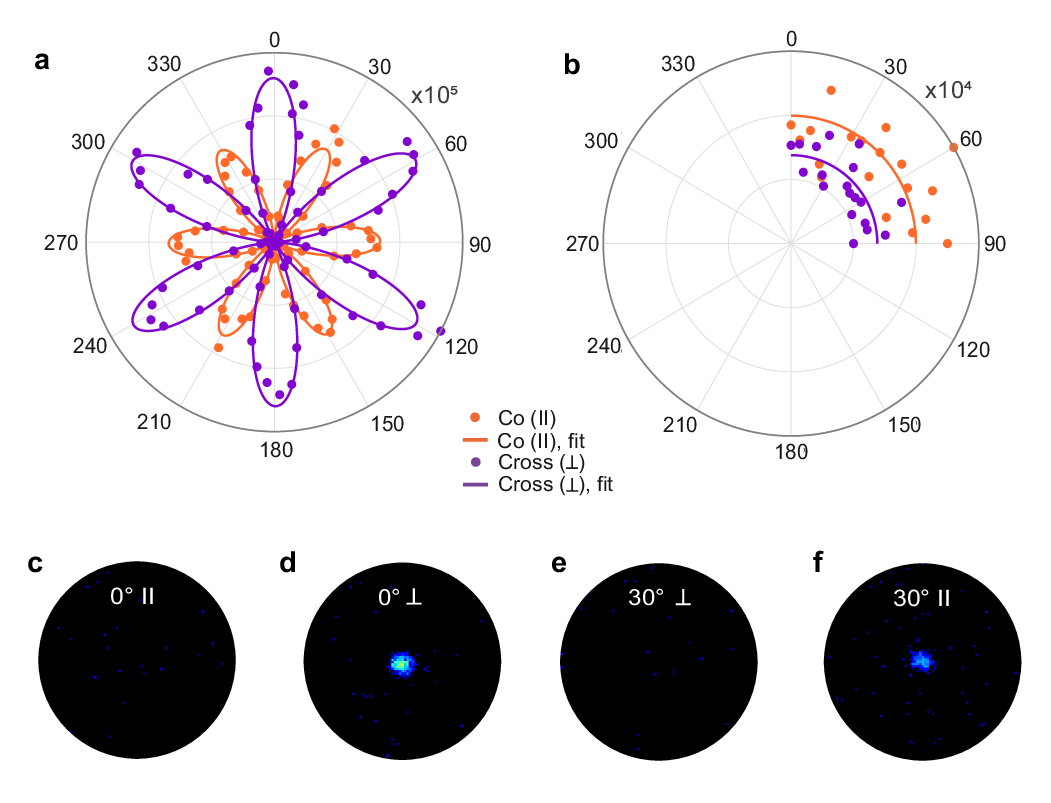}
 \caption{\textbf{(a)} Polarization-dependent intensity of the second harmonic for the monocrystalline Cu microflake at 2.5 mW/$\mu$m$^2$ excitation.Cross-polarized emission (purple), i.e. signals emitted perpendicular to the fundamental polarization, and co-polarized emission (orange), which is parallel to the fundamental polarization are shown. The data were fitted using the model by Sipe et al\cite{sipe_phenomenological_1987}. \textbf{(b)} Polarization-dependent intensity measured for the polycrystalline Cu film measured using the same experimental configuration. The response of the polycrystalline film is represented by the average of all data points.\textbf{(c-f)} Fourier images of the SHG recorded in co-polarized and cross-polarized detection for crystal orientations $\varphi = 0^\circ$ and $\varphi = 30^\circ$, using a 60$\times$ objective (NA = 0.7, acceptance angle = 97$^\circ$): \textbf{(c)} $\varphi = 0^\circ$, co-polarized; \textbf{(d)} $\varphi = 0^\circ$, cross-polarized; \textbf{(e)} $\varphi = 30^\circ$, co-polarized; \textbf{(f)} $\varphi = 30^\circ$, cross-polarized. No significant SHG signal was detected in panels (c) and (f). The bright central feature in (d) confirms dominant SHG emission in the cross-polarized channel at $\varphi = 0^\circ$, attributed to strong in-plane anisotropic nonlinear polarization components. When the crystal orientation is changed to $\varphi = 30^\circ$, the dominant SHG signal appears in the co-polarized channel (e). In both cases, the SHG emission is concentrated near the center of the Fourier plane, indicating that the emission is predominantly directed normal to the surface of the flake.}
\label{fig_polarized}
\end{figure}

To understand the fundamental nature of the SHG signal, we performed polarization-dependent measurements on monocrystalline Cu microflakes and their polycrystalline Cu counterparts, for which the bright-field micrographs were shown in Figure \ref{fig_flakefilm}a,c. In these measurements, we fixed the incident beam polarization and rotated the sample while recording the cross and co-polarized SHG intensities. Further details of these measurements are included in the Methods \ref{sec:Methods} section and the resulting plots are shown in Figure \ref{fig_polarized}. 

The six-petal SHG pattern in Fig.~\ref{fig_polarized}a confirms the anisotropic surface response of the Cu microflake and indicates that its surface symmetry corresponds to the point group \( C_{3v} \)~\cite{wu_neumanns_2022,bloch_detection_1992}, characterized by three-fold rotational symmetry. This behavior is consistent with observations reported for Au(111) flakes, where a similar SHG angular dependence was attributed to symmetry-allowed anisotropic tensor contributions on fcc (111) metal surfaces~\cite{boroviks_anisotropic_2021}. The SHG signal exhibits clear angular modulation in both co- and cross-polarized detection channels as the excitation polarization is rotated relative to the crystal axis. Different orientations probe distinct combinations of nonlinear susceptibility tensor elements, for instance the co-polarized signal peaks near \( 30^{\circ} \), where the excitation field couples most strongly onto tensor elements such as \( \chi^{(2)}_{xxx} \), corresponding to the dominant \( \sin(3\theta) \) terms in the fit and reflecting in-plane electric field coupling at the surface~\cite{lupke_resonant_1994}. In contrast, the cross-polarized signal is strongest around \( 0^{\circ} \), where the excitation lies between principal crystal axes, enhancing contributions from off-diagonal elements such as \( \chi^{(2)}_{xxy} \), which couple orthogonal excitation and emission fields.

In the case of the polycrystalline Cu film, shown in the polar plot of Fig.~\ref{fig_polarized}b, the absence of angular dependence is consistent with expectations, resulting from the averaging of randomly oriented grains under excitation by a beam with a spot size of approximately \( 2~\mu\text{m} \)~\cite{krause_optical_2004}.

Further insight on the SHG signal from Cu monocrystalline flake is provided by the Fourier plane images shown in Figure \ref{fig_polarized} c-f, which reveal the angular distribution of the SHG emission. These measurements were performed on a different Cu microflake and the corresponding polar plot is included in Figure S2, in good agreement with the polar plots in \ref{fig_polarized}. The bright central spot observed in the cross-polarized channel at 0° (panel d), and similarly in the co-polarized channel at 30° (panel e), confirms that the SHG emission is directed predominantly along the surface normal (out-of-plane), while no significant SHG signal was detected in panels (c) and (f). This symmetry-driven directionality could be exploited in future nonlinear optical devices that require polarization control or angular filtering of SHG signals.



\section{Conclusions} \label{sec:Conclusions}

In summary, we present a comprehensive study of the anisotropic nonlinear response of monocrystalline Cu with (111) surface orientation at ambient conditions, thanks to the distinct stability of the Cu microflakes under illumination. We confirm that the SHG measurements can determine the crystal structure, which for monocrystalline Cu shows point group C\textsubscript{3v} characterized by threefold rotational symmetry, unlike a polycrystalline film. Furthermore, we observe a pronounced cross-polarized SHG emission from Cu, bridging the gap between the nonlinear optical behavior of Cu and Au microflakes as revealed by polarization-dependent measurements. 
Overall, we envision these findings will promote new applications and research directions in the development of copper-based nonlinear plasmonic and nanophotonic devices, as well as foster better integration of SHG spectroscopy as a rapid characterization tool in energy conversion and photocatalysis.



\newpage
\bibliographystyle{achemso}
\providecommand{\latin}[1]{#1}
\makeatletter
\providecommand{\doi}
  {\begingroup\let\do\@makeother\dospecials
  \catcode`\{=1 \catcode`\}=2 \doi@aux}
\providecommand{\doi@aux}[1]{\endgroup\texttt{#1}}
\makeatother
\providecommand*\mcitethebibliography{\thebibliography}
\csname @ifundefined\endcsname{endmcitethebibliography}  {\let\endmcitethebibliography\endthebibliography}{}

 \section*{Data Availability Statement} \label{sec:data}
All the data supporting the findings of this study will be available at Zenodo.
\section*{Supporting Information} \label{sec:supportinginfo} 

Schematic of the setup used for SHG measurements, polarization-resolved polar plot. \\
\section*{Acknowledgements} \label{sec:acknowledgements}
 E.N.D.  and  G.T. acknowledge the SNSF Eccellenza  Grant PCEGP2-194181. D.P. acknowledges the support of Laidlaw Internship Program. A.R.B. acknowledges support of SNSF Eccellenza Grant PCEGP2-194181 and SNSF Swiss Postdoctoral Fellowship TMPFP2\_217040. We would like to acknowledge Dr. Priscila Vensaus for the assistance with AFM measurements and Mrs. Diana Dall'Aglio for the assistance and discussion on the setup and BFP measurements.
\section*{Author contributions} \label{sec:contributions}
E.N.D. prepared the samples with help from D.P.. E.N.D. and A.R.B. performed the experiments with help from
O.C.K.. E.N.D. and A.R.B. analyzed and interpreted the data. E.N.D. wrote
the paper, with input from all the other authors. G.T. supervised the project. 
\section*{Competing interests} \label{sec:comp}
The Authors declare no competing interests.


\newpage
\section*{Table of Content Graphic} \label{sec:toc}

\begin{figure}[H]
    \centering  
\includegraphics[width=12cm]{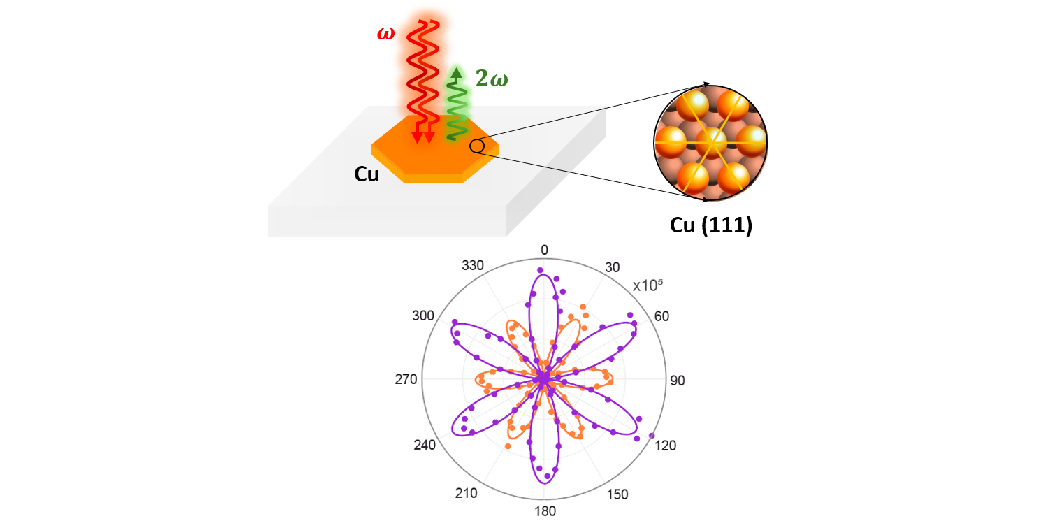}
\label{fig_toc}
\end{figure}

\end{document}